\documentclass[11pt]{article}
\usepackage{moriond,epsfig}

\newcommand{\dzero}     {D\O}

\newcommand{\rargap}    {\mbox{ $\rightarrow$ }}
\newcommand{\ttbar}     {\mbox{$t\bar{t}$}}
\newcommand{\ppbar}     {\mbox{$p\bar{p}$}}

\def\be{\begin{equation}}
\def\ee{\end{equation}}
\def\bea{\begin{eqnarray}}
\def\eea{\end{eqnarray}}

\begin{document}
\vspace*{4cm}
\title{Evidence for Single Top Production at the Tevatron}

\author{Supriya Jain}

\address{Homer L. Dodge Department of Physics and Astronomy, University of Oklahoma,\\
440 W. Brooks St., Norman, Oklahoma 73019, USA\\
(On behalf of D0 and CDF Collaborations)}
\maketitle
\vspace*{-0.2in}
\abstracts{
We present first evidence for the production of single top quarks at
the Fermilab Tevatron {\ppbar} collider. Both D0 and CDF experiments
have measured the single top production cross section with a 
3-standard-deviation significance using 0.9~fb$^{-1}$ and 2.2~fb$^{-1}$ of 
lepton+jets data, respectively. A direct measurement of the CKM matrix element that
describes the $Wtb$ coupling is also performed for the first time.
}

\vspace*{-0.35in}
\section{Introduction}
\label{sec:intro}
\begin{figure}[!h!tbp]
\begin{center}
\psfig{figure=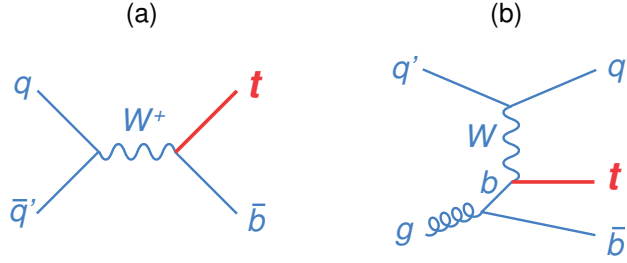,height=1.5in}
\caption{Representative Feynman diagrams for (a) s-channel
and (b) t-channel single top quark production.
\label{fig:feynman}}
\end{center}
\end{figure}

The top quark was discovered in 
1995 at the Tevatron~\cite{top-obs-1995} in pair production mode 
($t \bar{t}$ events) involving strong interactions. But the 
standard model (SM) also predicts the production of single 
top quarks through the electroweak exchange of a $W$ boson. 
At the Tevatron, the $s$- and $t$-channel production modes 
illustrated in Fig.~\ref{fig:feynman} are dominant. An 
observation of 
single top production is interesting since it would provide a 
direct measurement of the CKM matrix element 
$\left| V_{tb} \right|$ and top quark polarization, 
in addition to probing possible new physics in the top 
quark sector. But observing single top quarks is challenging 
on account of smaller cross sections (about half that of 
$t \bar{t}$ production) as well as larger backgrounds on 
account of only one massive object in the final state. The 
total predicted cross section for single top quarks at the 
Tevatron is $2.86\pm 0.33$~pb~\cite{singletop-xsec-sullivan}
for a top quark mass of 175 GeV. 
Simple selections that exploit kinematic features
of the signal final state are not sufficient, and one needs 
sophisticated multivariate techniques to 
separate single top events from the overwhelming backgrounds. 
Both D0 and CDF have performed several multivariate 
analyses~\cite{cdf-singletop,d0-singletop} 
to identify the putative signal in data, and we 
report here the results. 

\vspace*{-0.1in}
\section{Event Selections}
\label{sec:selections}
The experimental signal for single top quark events consists of one
isolated high transverse momentum, central pseudorapidity
charged lepton (electron $e$ or muon $\mu$) 
and missing transverse energy
from the decay of a $W$~boson from the top quark decay,
accompanied by a $b$~jet also from the top quark decay. There is always a
second jet, which originates from a $b$~quark produced with the top
quark in the s-channel, or which comes from a forward-traveling up- or
down-type quark in t-channel events. Some t-channel events have a
detectable $b$~jet from the gluon splitting to $b\bar{b}$. 
Both D0 and CDF apply selections to keep signal-like events while 
rejecting backgrounds. 
CDF includes events with two or three jets with 
single or double $b$-tags, while D0 includes four-jet events also 
in which the additional jet is from initial-state or final-state radiation. 

\vspace*{-0.1in}
\section{Signal and Background Modeling}
\label{sec:model}
Modeling of signals and backgrounds is done using both 
Monte Carlo simulations and data. 
The main backgrounds
are: $W$+jets; $t \bar{t}$ in
the lepton+jets and dilepton final states, when a jet or a lepton is
not reconstructed; and multijet production, where a jet is
misreconstructed as an electron or a heavy-flavor quark decays 
to a muon that is misidentified as isolated from the jet. 
The $W$+jets includes both $W$+light-flavor jets ($Wjj$) and 
$W$+heavy-flavor jets ($Wb\bar{b}$, $Wc\bar{c}$, and $Wcj$). 
There is also a small contribution from diboson and $Z$+jets events. 
D0 analysed 0.9~fb$^{-1}$ of lepton+jets data, while CDF analyzed 
2.2~fb$^{-1}$ of data. The expected and observed event yields from 
both experiments after all selections are shown in 
Table~\ref{tab:event-yields}. 
%
\begin{table}[!h!tbp]
\caption[eventyields]{Event yields, (left) at D0, 0.9~fb$^{-1}$, and (right) at CDF, 2.2~fb$^{-1}$, for $e$ and $\mu$, 1 $b$~tag and 2 $b$~tag 
channels summed. The diboson and $Z$+jets events are 
included in the overall $W$+jets background at D0.}
\label{tab:event-yields}
\begin{minipage}[b]{0.5\linewidth}
\centering
\begin{tabular}{l@{\extracolsep{\fill}}r@{\extracolsep{0pt}$\pm$}l@{}%
                 @{\extracolsep{\fill}}r@{\extracolsep{0pt}$\pm$}l@{}%
                 @{\extracolsep{\fill}}r@{\extracolsep{0pt}$\pm$}l@{}}
Source           & \multicolumn{2}{c}{2 jets}
                 & \multicolumn{2}{c}{3 jets}
                 & \multicolumn{2}{c}{4 jets} \\
\hline	     
$tb$                 &  16  &   3  &   8  &  2  &   2  &  1  \\
$tqb$               &  20  &   4  &  12  &  3  &   4  &  1  \\
\hline                                                              
\ttbar               &  59  &   10  &  135  &  26  &  154  &  33  \\
$Wb\bar{b}$  & 261  &  55  & 120  & 24  &  35  &  7  \\
$Wc\bar{c}$,$Wcj$  & 151  &  31  &  85  & 17  &  23  &  5  \\
$Wjj$              & 119  &  25  &  43  &  9  &  12  &  2  \\
Multijets         &  95  &  19  &  77  & 15  &  29  &  6  \\
\hline                                                              
Total background~~  & 686  &  41  & 460  & 39  & 253  & 38  \\
Data             & \multicolumn{2}{c}{697}
                    & \multicolumn{2}{c}{455}
                    & \multicolumn{2}{c}{246}
\end{tabular}
\end{minipage}
\hspace{0.5cm}
\begin{minipage}[b]{0.5\linewidth}
\centering
\vspace*{0.2cm}

\begin{tabular}{l@{\extracolsep{\fill}}r@{\extracolsep{0pt}$\pm$}l@{}%
                 @{\extracolsep{\fill}}r@{\extracolsep{0pt}$\pm$}l@{}}
Source           & \multicolumn{2}{c}{2 jets}
                 & \multicolumn{2}{c}{3 jets}\\
\hline	     
$tb$                 &  41.2  &   5.9  &   13.5  &  1.9  \\
$tqb$               &  62.1  &   9.1  &   18.3  &  2.7  \\
\hline                                                              
\ttbar               &  146.0  &   20.9  &  338.7  &  48.2  \\
$Wb\bar{b}$  & 461.6  &  139.7  & ~~141.1  & 42.6   \\
$Wc\bar{c}$,$Wcj$  & 395.0  &  121.8  &~~  108.8  & 33.5   \\
$Wjj$              & 339.8  &  56.1  &  101.8  &  16.9    \\
Multijets         &  59.5  &  23.8  &  21.3  & 8.5   \\
Diboson         & 63.2   &  6.3  &  21.5  & 2.2   \\
$Z$+jets         & 26.7   & 3.9   & 11.0   & 1.6   \\
\hline                                                              
Total background~~  & 1491.8  &  268.6  &~~ 754.8  & 91.3   \\
Data             & \multicolumn{2}{c}{1535}
                    & \multicolumn{2}{c}{712}
\end{tabular}
\end{minipage}
\end{table}
%

%
\section{Cross Section Measurements at D0}
\label{sec:d0}
\begin{figure}[!h!tbp]
\psfig{figure=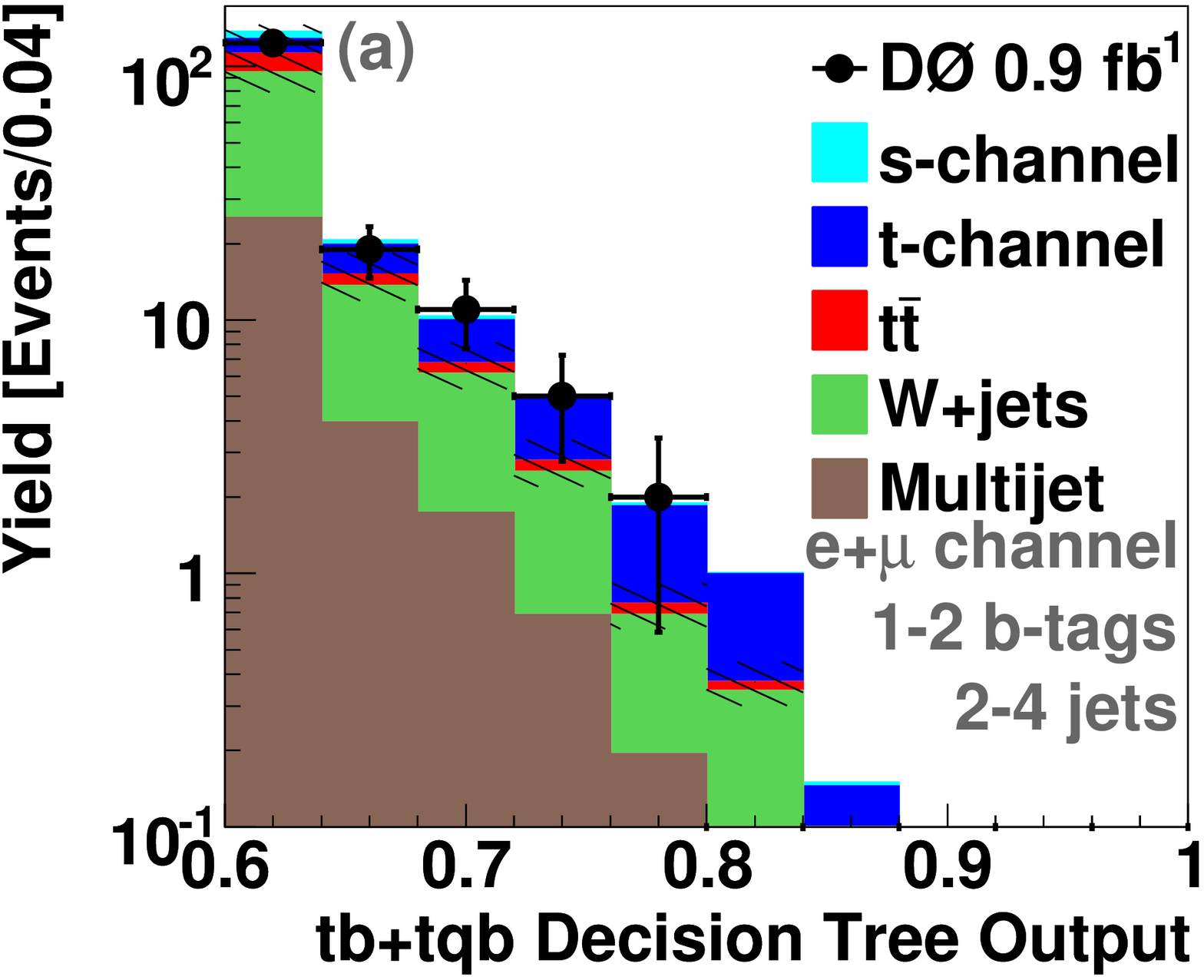,height=1.5in}
\psfig{figure=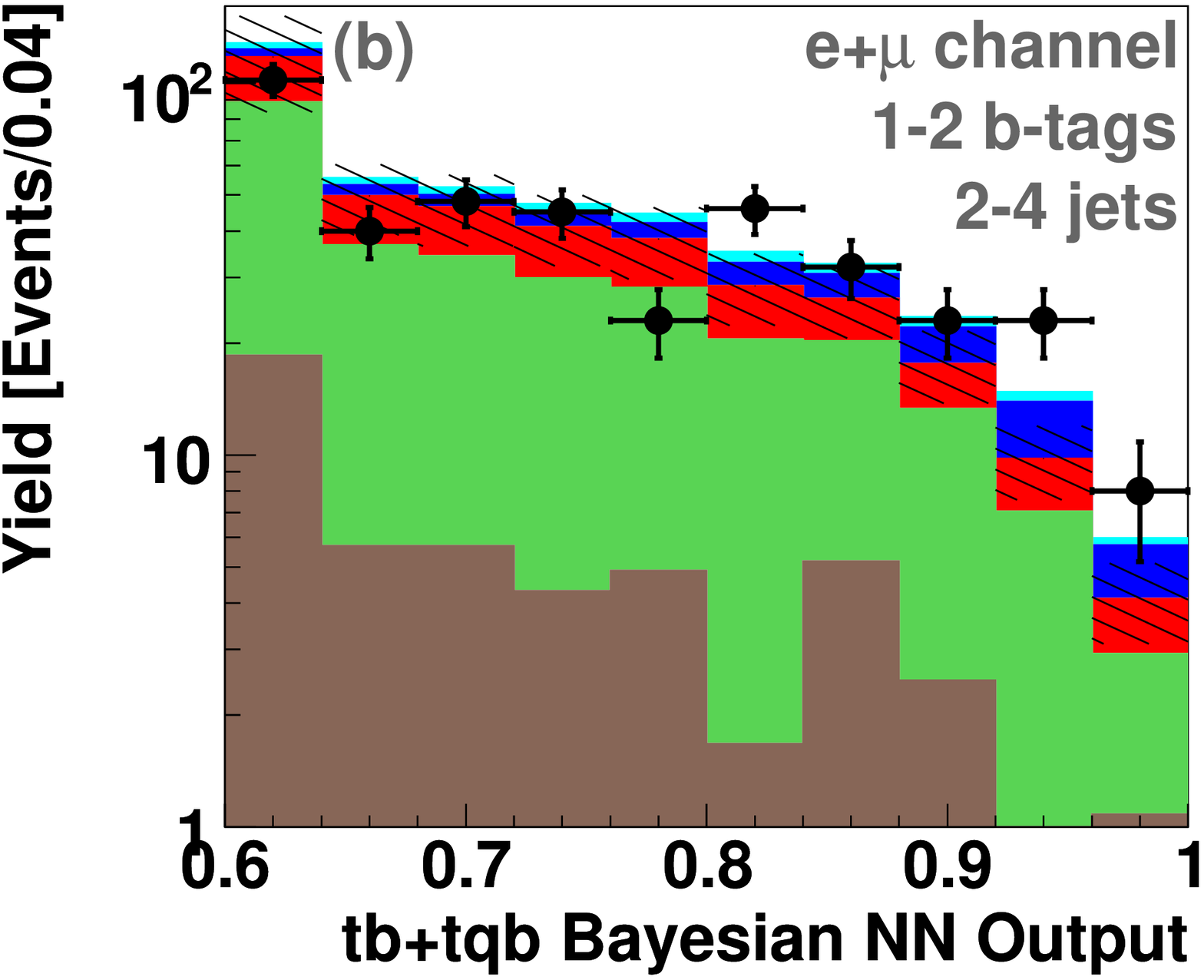,height=1.5in}
\psfig{figure=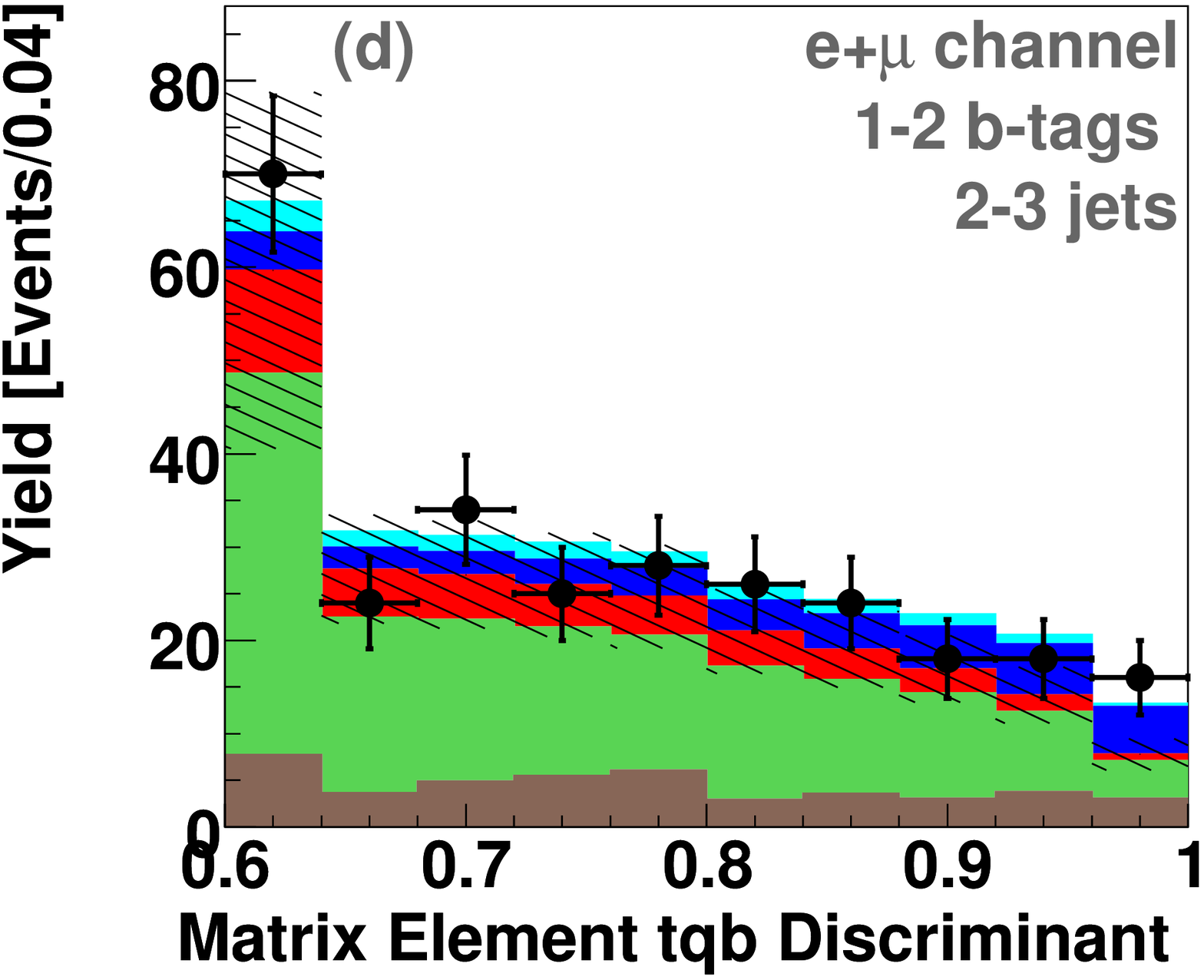,height=1.5in}
\caption{The signal regions of the multivariate 
discriminants at D0: (a) DT, (b) BNN, and (c) ME.
\label{fig:disc-d0}}
\end{figure}

\begin{figure}[!h!tbp]
\begin{center}
\vspace{-0.2in}
\psfig{figure=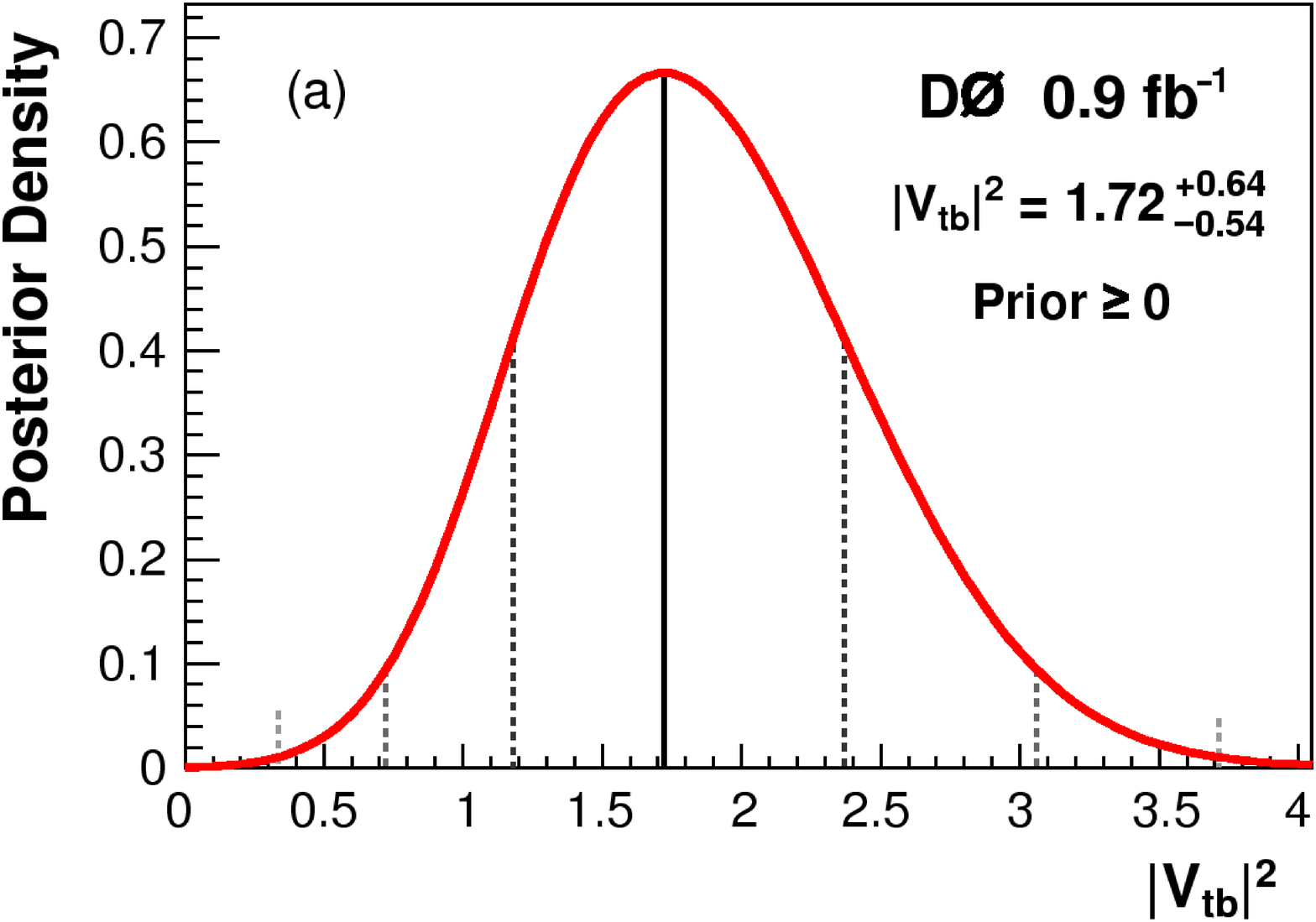,height=1.5in}
\psfig{figure=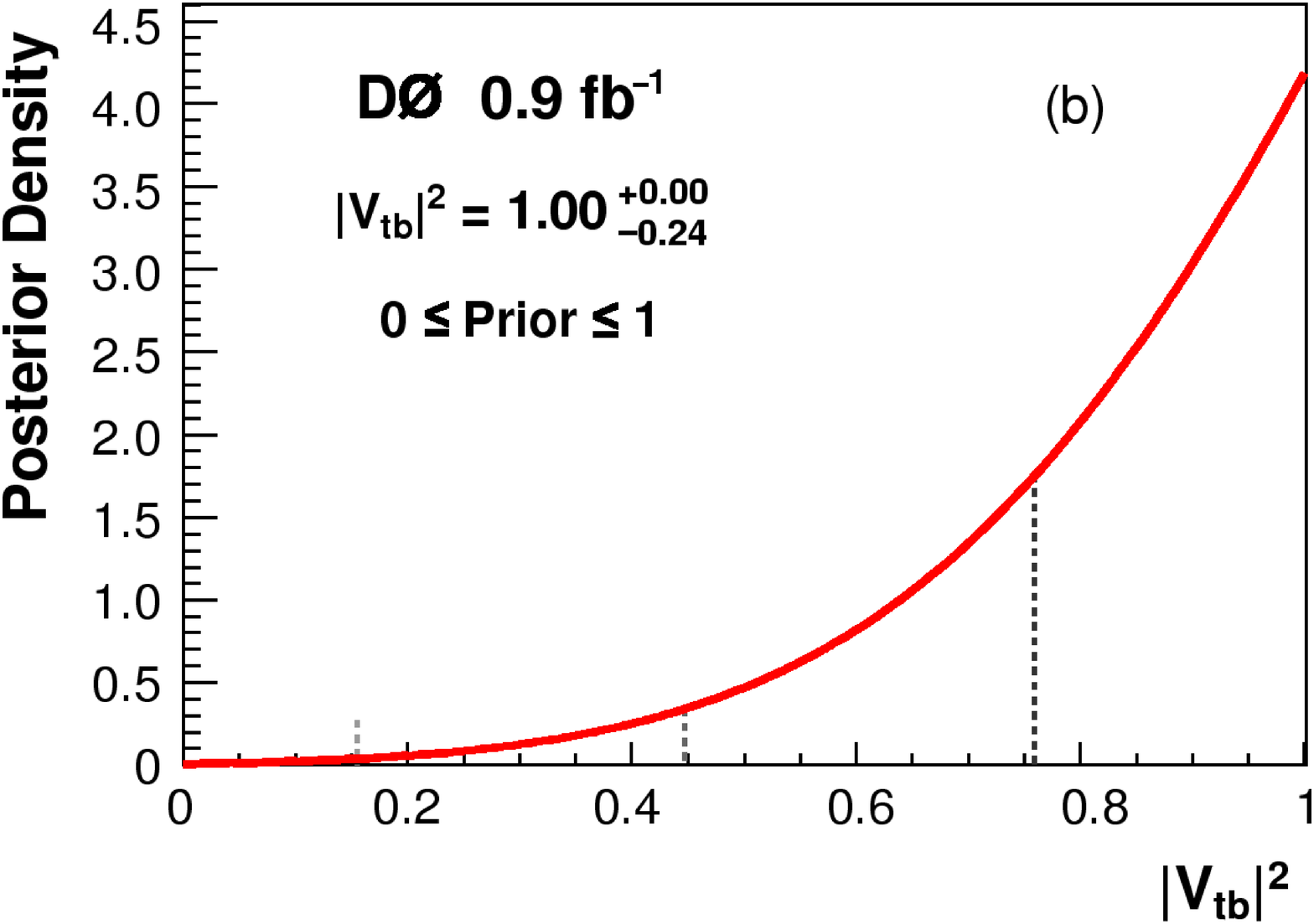,height=1.5in}
\caption{The Bayesian posterior density distributions for $|V_{tb}|^2$
for (a) prior$\geq$0, and (b) 0$\leq$prior$\leq$1 at D0. 
The dashed lines show the positions of the one, two, and three standard 
deviations away from the peak of each curve.
\label{fig:vtb-d0}}
\end{center}
\vspace{-0.2in}
\end{figure}

Three different multivariate analyses were performed at D0 
in order to separate the single top signal from backgrounds: 
boosted decision trees (DT), 
Bayesian neural networks (BNN), and matrix elements 
(ME). The discriminants from 
each analysis are shown in Fig.~\ref{fig:disc-d0}. 
Cross section measurements are 
extracted from a binned likelihood of the discriminants 
separately for each analysis. 
The results are then combined using the BLUE 
(best linear unbiased estimate) method 
yielding 
{\bf $\sigma({\ppbar}{\rargap}tb+X,~tqb+X) = 4.7 \pm 1.3$~pb}. 
A large  ensemble of pseudo-datasets is created 
to estimate the significance of measurements. It is:
$$
\begin{array}{lll}
\rm Expected~p-value: & 1.1\% & (2.3~\rm standard~deviations)\\
\rm Observed~p-value: & 0.014\% & (3.6~\rm standard~deviations).
\end{array}
$$
Additionally, D0 also measured the value 
of $|V_{tb}|$ using two different assumptions for the prior probablity 
density of $|V_{tb}|^2$ as shown in Fig~\ref{fig:vtb-d0}.
%

\section{Cross Section Measurements at CDF}
\label{sec:cdf}
\begin{figure}[!h!tbp]
\begin{center}
\vspace{-0.2in}
\psfig{figure=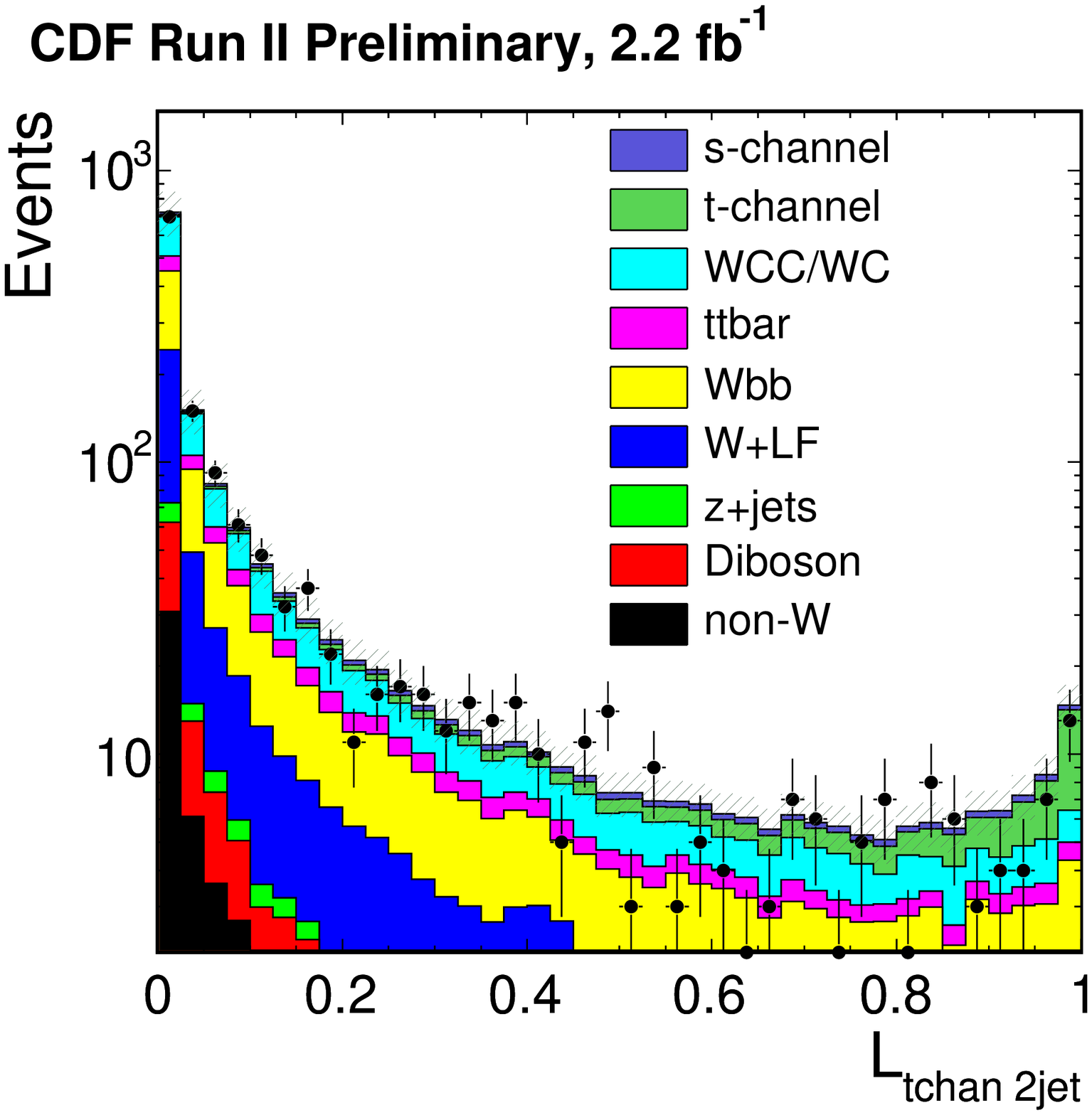,height=1.5in}
\psfig{figure=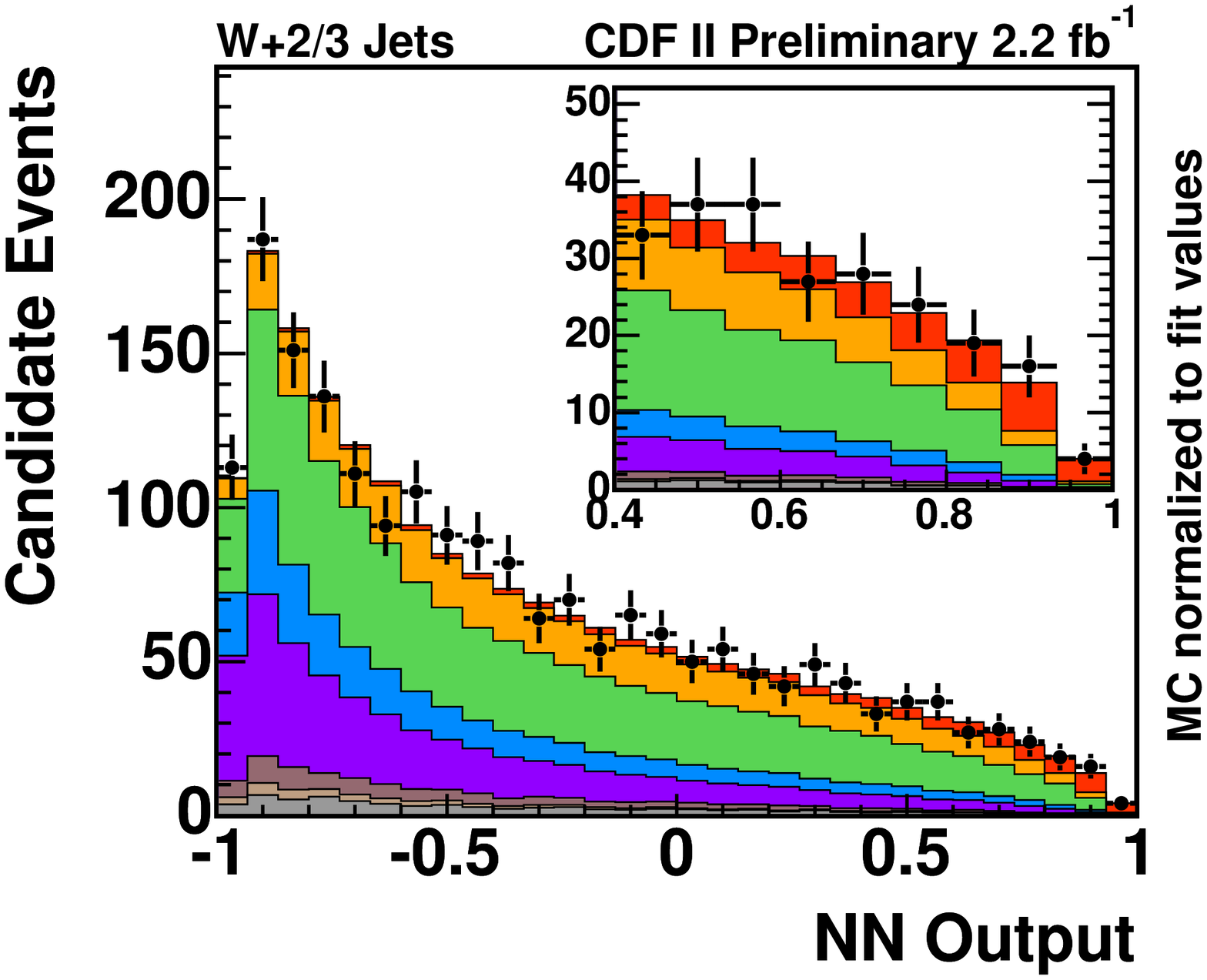,height=1.5in}
\psfig{figure=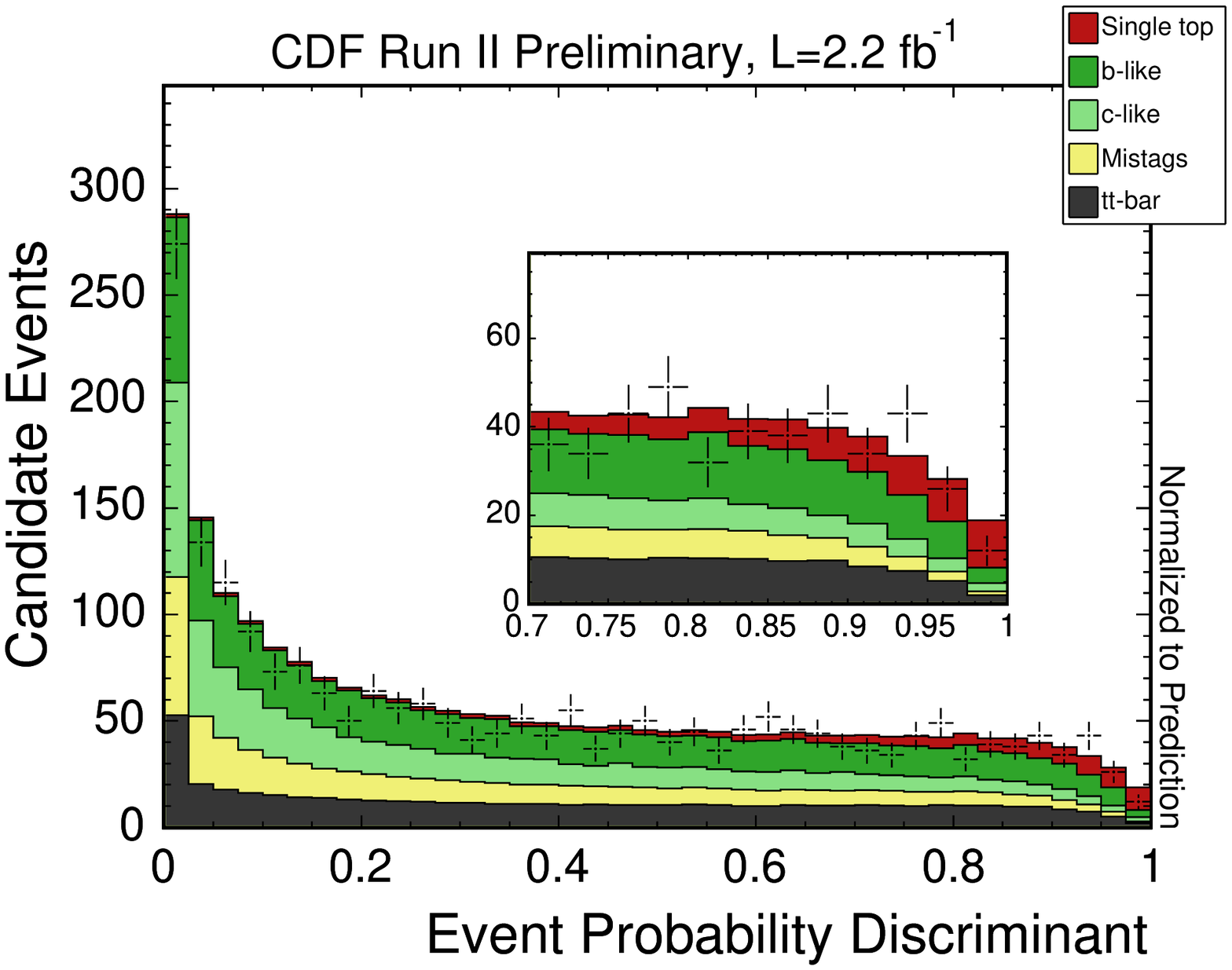,height=1.5in}
\caption{Multivariate discriminant output distributions 
at CDF: LF (left), NN (middle), and ME (right).
\label{fig:disc-cdf}}
\end{center}
\end{figure}

CDF performed the following multivariate analyses: 
likelihood function (LF), 
neural networks (NN), and matrix elements 
(ME). The discriminants from 
each analysis are shown in Fig.~\ref{fig:disc-cdf}. 
Subsequent cross section measurements using a binned likelihood 
are the following:
{
$$
\begin{array}{lll}
\bf
\sigma\left({\ppbar}{\rargap}tb+X,~tqb+X\right)
& = 1.8 ^{+0.9}_{-0.8}~{\rm pb} & {\rm (Likelihood~function)}\\
& = 2.0 ^{+0.9}_{-0.8}~{\rm pb} & {\rm (Neural~networks)}\\
& = 2.2 ^{+0.8}_{-0.7}~{\rm pb} & {\rm (Matrix~elements)}.
\end{array}
$$
}
From the above results, $|V_{tb}|$ is measured to be $0.78 ^{+0.18}_{-0.21}$, and a lower 
limit of 0.41 is also set at 95\% confidence level (CL). 

\vspace*{-0.1in}
\section{Single Top Projections}
\label{sec:projections}
\begin{figure}[!h!tbp]
\begin{center}
\psfig{figure=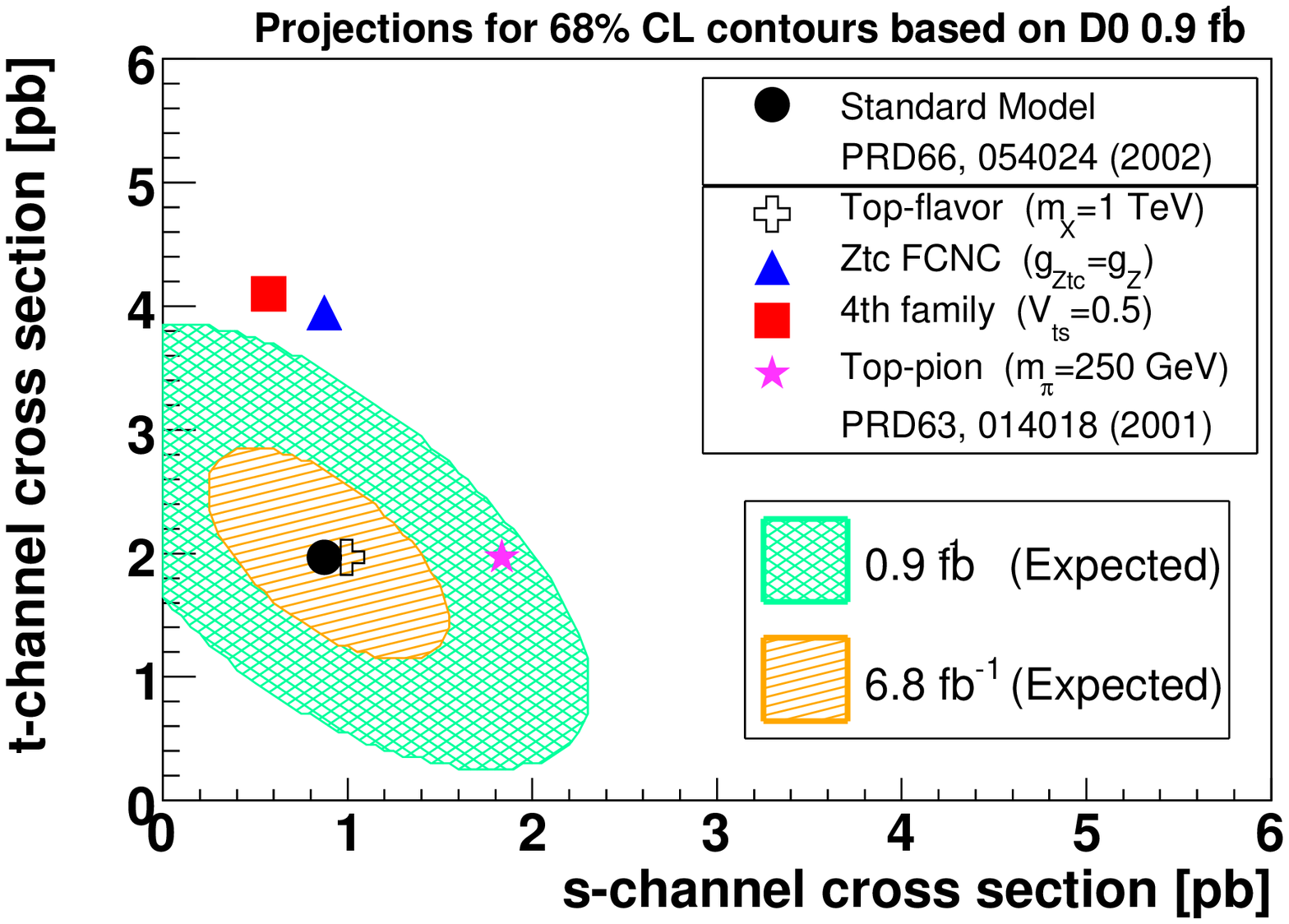,height=1.4in}
\psfig{figure=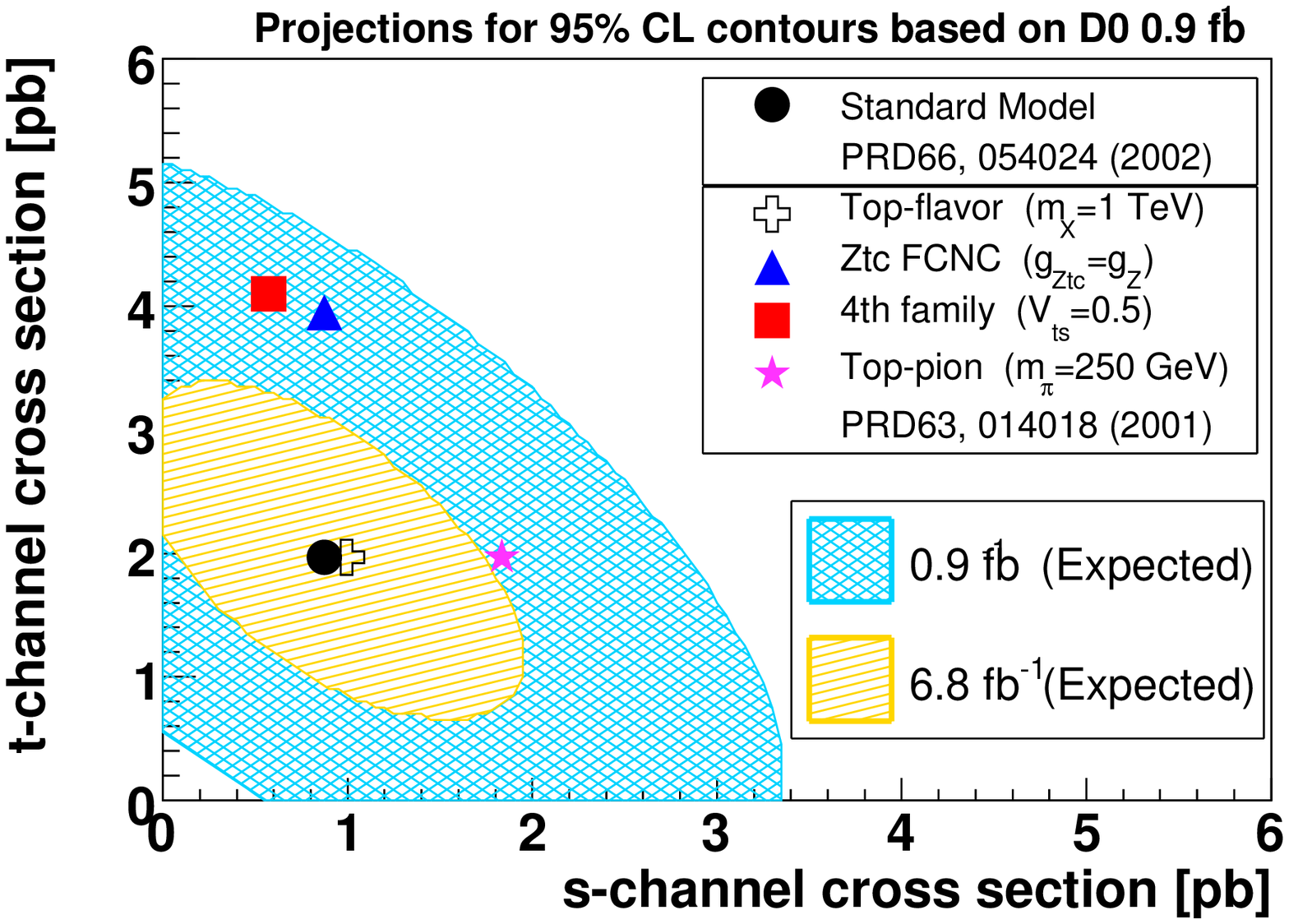,height=1.4in}
\psfig{figure=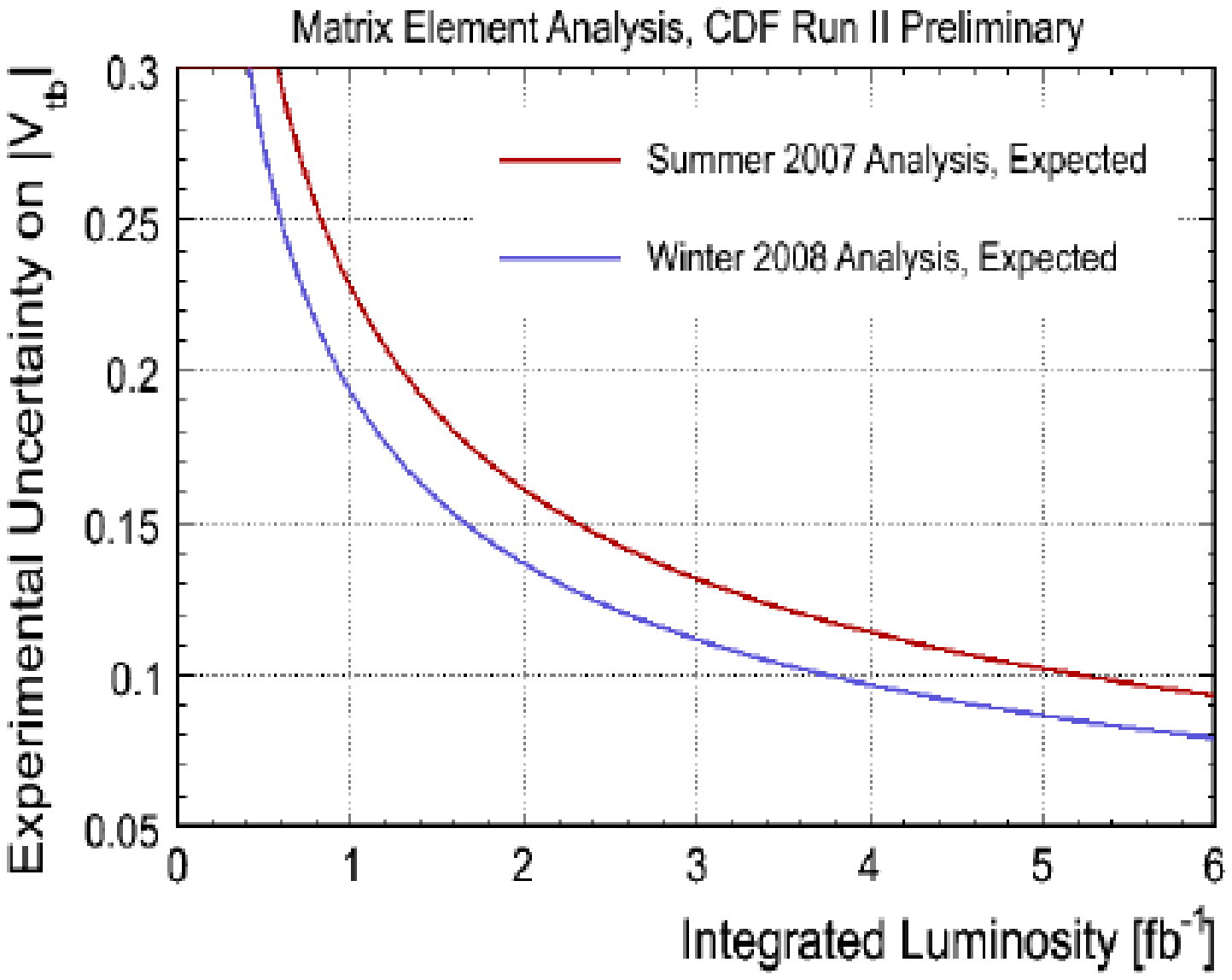,height=1.5in}
\caption{The allowed contours at 68\% and 95\% CL in a 
two-dimensional plane of $\sigma_s$ versus $\sigma_t$, and the uncertainty on 
$|V_{tb}|$ expected at different values of 
integrated luminosity based on D0 and CDF data.
\label{fig:proj}}
\end{center}
\vspace{-0.2in}
\end{figure}

D0 and CDF also made projections for different measurements based on their current analyses. 
The allowed contours at 68\% and 95\% CL in a 
two-dimensional plane of $\sigma_s$ versus $\sigma_t$, and the uncertainty on 
$|V_{tb}|$, expected at different values of 
integrated luminosity are shown in Fig.~\ref{fig:proj}. 
We see that it is possible to exclude several models beyond SM at about 7~fb$^{-1}$, and 
$|V_{tb}|$ can be measured with a precision better than 10\% as the Tevatron collects more data. 
%

\section{Summary}
\label{sec:summary}
To conclude, searches for single top quarks have been performed 
at the Tevatron using multivariate techniques to separate the 
signal from the huge backgrounds. Both D0 and CDF 
have measured the single top production cross section with a 
3-standard-deviation significance using 0.9~fb$^{-1}$ and 2.2~fb$^{-1}$ of 
lepton+jets data, respectively. A direct measurement of the CKM matrix 
element $|V_{tb}|$ is also performed for the first time. As the Tevatron 
collects more data, both experiments continue to improve their analyses in order to increase the sensitivity to a 
possible observation. 
%

\section*{Acknowledgments}
The author would like to thank the organizers for creating a 
fruitful collaborative environment at the conference,
and also the European Union for their support under 
the Marie Curie programme.


\end{document}